# Comment on ArXiv: hep-ph/0902.3613 by P. Bicudo et al.

Recently few authors present the paper devoted to the Δ- resonance spectra and new quartet scheme were developed there [1]. The authors also generated interesting scheme for *j*-, *k*-, and mass-scaling of appropriate spectra and discover the possibility of extracting the running quark mass from the data. The purpose of this short Comment is to quantify some of the claims [1] and clarify many results and future developments which are possible.

On a page 2, [1] we find - "middle and high range of the spectrum of light hadrons, extending between about 1 GeV and 3 GeV". In fact, we can see from the PDG08 data that $N$, $\Delta$ -spectrum is extending up to 4.1 GeV, and someone has to analyze this entire region, 1 – 4.1 GeV.

The authors [1] claim and prove extensively throughout the paper that- " three-quark states naturally group into the quartets with two states of each parity". This is an interesting observation, but so far nothing similar has been observed in any experiment.

On a page 2, [1] we find: "we study family of maximum-spin excitations $\Delta^*$, the leading Regge trajectory of the Δ baryon spectrum" – But, in fact, the authors studied the sequence of the states with $J$=3/2, 5/2, 7/2, 9/2, 11/2, 13/2…, which is not a leading Δ Regge trajectory (RT), which is $\Delta(3/2)^+$ - $\Delta(7/2)^+$ - $\Delta(11/2)^+$ - $\Delta(15/2)^+$. I'm not sure that all the members of their Δ-sequence [1] actually correspond to the maximum spin s=3/2 of the Δ.

Another interesting estimate from the [1] is: "We advocate that a measurement of masses of high-partial wave Δ-resonances with an accuracy of 50 MeV should be sufficient to unambiguously establish the approximate degeneracy, and test the concept of running quark mass in the infrared". – How this estimate of the 50 MeV accuracy has been done? Someone has to take into account the partial- and the full-width of each resonance, the mass value, and even the lineshape. This estimate also should correlate with the theory itself, which predicts a quartets, and values of a splittings for each $\Delta^*$. So, this estimate could be shifted either way, and the authors never used it throughout [1] anyways.

On a page 4 the authors [1] are trying to justify their expansion procedure in $m(k)/k$ series, claiming this ratio as a small parameter. But, as one can see from their Fig.1, such a parameter is an order of unity in the low-$k$ region of the $\Delta(3/2)^+$. So, if such an expansion fails in one region, how can we expect it to work for the whole $k$-space? Many physical observables will involve integration over the whole $d^3k$ space, and considerable errors will accumulate.

On a page 5 Bicudo et al. [1] push very hard to define the quartet states and the quartet basis for Δ-baryons. But we have to admit – these states and basis are valid only for high-lying resonances, with $k/m \gg 1$, and the states $\Delta(3/2)$ and $\Delta(5/2)$ would not qualify. We agree that an idea is interesting, but it's has not been properly justified, and the whole notion of quartets and it's relation to the current data is doubtful.

The same arguments were proposed on page 6 [1], but it's clear that the whole "quartet basis" procedure will work only for really high $\Delta^*$. Surprisingly, the authors [1] didn't estimate numerically, from what $J$-value for $\Delta^*$ this ansatz will really hold.

It is really interesting to establish $j$-scaling for the $|M^+ - M^-|$ and relate it to the $k$-scaling of the running quark mass. But then, the authors [1] claim on a page 6: "Naturally, the two approximately degenerate masses $M^+$ and $M^-$ are both part of the same leading Regge trajectory, phenomenologically fixing their j-scaling"

$$j = \alpha_0 + \alpha(M^{+-})^2 \underset{j \to \infty}{\to} \alpha(M^{+-})^2 \qquad (1)$$

The Eq.(1) will work only at asymptotic $j$-values, as we proved in [2-3], and not in the current experimental resonance region. Again, the series of $\Delta^*$, shown at Fig.2, is not a leading Regge trajectory (RT).

The authors [1] wrote on page 5 - "Large $j$ is equivalent to large quark orbital angular momentum $l$ since the spin is finite, and also to a large average linear momentum $<k>$ ." This is a very delicate matter of judgment, and it would be terrific to find a realistic relations between $j$, $k$, and $l$ in all kinematic domain.

As one can see, the wave functions at Fig.2 does not seem to be normalized at $k=0$, and so, the mean value $<k>$ could be very different from that of the Fig.2. The authors [1] should provide reader with $<k>_i$ value for each $\Delta^*$ RT, so one can judge the major claim about high $<k>_i$ for all RT's. We should ask, whether Bicudo et al. [1] made a fits for the full $N^*$ - $\Delta^*$- spectra? It is necessary, and major model parameters should be fixed this way. There is a discrepancy between the lattice quark running masses in Fig.1 and Fig.2. In Fig.1 for high-$k$ region lattice results in Coulomb and Landau gauges are very different from ~ 3 GeV, where Coulomb quark mass reach ~ 25 MeV, while the Landau quark mass ~ 80 MeV. On Fig.2 we can see only one lattice curve for the quark mass, which reaches ~75 MeV at $k \approx 6$ GeV. So, why the authors have done Figs.1 and 2 differently, because this high-$k$ region is crucial for the quartet scheme?

The authors [1] make analogies on page 7, between Eq.(9):

$$|M^+ - M^-| \to c_3 m(<k>) <H_\chi^{QCD'}> / <k> \qquad (2)$$

and the Gell-Mann-Oakes-Renner relation, Eq.(10)

$$M_\pi^2 = - m_q <\psi bar \psi> / f_\pi^2 \qquad (3)$$

I don't think, they are analogous, because there are very different physical objects involved.

On a page 8 the authors [1] computed *j*-scaling for the matrix element $<H_\chi^{QCD'}>$. I'm not sure of their factorization technique, and the results for the *j*-scaling. In our NRQM, which used HHF method [2,3], one will always get an *infinite system* of the radial equations, connected due to the angular wave functions–hyperspherical harmonics. So, there is no possibility to use or agree with such an ansatz,

$$< H_\chi^{QCD'}> \;\propto\; < H_\chi^{QCD'}>_{angular} \;*\; < H_\chi^{QCD'}>_{radial}$$

because it is simply wrong. The authors [1] note, in passing, that spin-independent potential scales like $1/j^{\,3/2}$, but we have also all kinds of spin-dependent parts in Hamiltonian, so how should correct *j*-scaling formula read? What is the meaning of number "i" in the exponents of Eqs.15, 16? Equations (15) and (16) for the running quark mass are quite different in *k*- and *j*-, so what can we learn from them?

The authors [1] consider that there is only *one* scale, relevant to this $\Delta$-spectrum, and it's related to the $\Lambda_{QCD}$-scale. But the situation for hadrons were discussed recently by few authors – Afonin [4], Ribeiro et al. [5], and it has been shown that there should exist a *few* scales and correspondingly a few fundamental parameters – $\Lambda_{QCD}$, $\Lambda_{BCS}$, $\Lambda_{CSB}$, $\Lambda_{CSR}$. Maybe an idea of only one scale [1] is wrong, and so we should deal with few and *different* scaling laws (see our Fig.1).

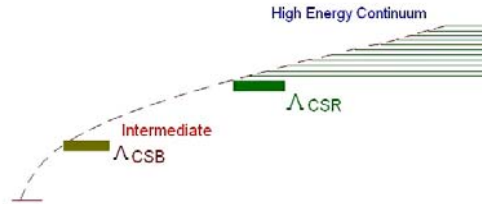

Fig.1 Different energy scales for the hadrons

Recently Afonin clearly shows for the light mesons [4] that "The physics above $\Lambda_{CSR}$ is indistinguishable from the physics of perturbative QCD continuum. The scale $\Lambda_{CSR}$ marks *transition* from intermediate to high energy like the scale $\Lambda_{CSB}$ does from low to the intermediate one. Thus, if experimentally confirmed, the scale $\Lambda_{rest}$ is of great importance because it marks the upper bound of resonance physics in the light quark sector of QCD" – so, we clearly see that Afonin implied the natural existence of the *broad intermediate* region with distinctive physics. In their analysis of the heavy-light mesons,

Ribeiro et al. [5] introduced clearly a few scales and even had estimates for the corresponding fundamental parameters:

$\Lambda_{CSR} \approx 10\Lambda_{BCS} \approx 2.5$ GeV, $\Lambda_{BCS} \approx 250$ MeV, corresponds to the light quark condensate, and $\Lambda_{BCS} \approx \Lambda_{QCD}$.

It is interesting, that recently Shifman also found out [6]: "Asymptotic chiral symmetry restoration might be possible if a nonlinearity (convergence) of the Regge trajectories in an "intermediate window" of *n*, *J*, beyond the explored domain, takes place. This would signal the failure of the quasiclassical picture." It is in accord with our findings – nonlinearity of the hadronic RT is playing together with chiral symmetry restoration [2-3].

In Chapter 5 the authors [1] explained the procedure for computing parity doublets. They have "neglected all the magnetic interactions, which makes *Δ*-nucleon mass splitting too small". But this will ruin any attempt for the *N-Δ* spectra fit. In other words, their parameters, corresponding to the Fig.3, would not fit *N-Δ* spectra, which is not fair. For example, we took σ from our model [2,3], which provide an excellent fit to the whole *N-Δ* spectra, up to the highest *J*, and it is $\sigma \approx 0.07$ GeV$^2$, compared to $\sigma \approx 0.135$ GeV$^2$ from [1]. Actually, it is too bad that the authors [1] didn't quote all their model parameters, and there is no Table with Δ-masses, which will help to understand the real quality of the fit. In Fig.3 the authors [1] didn't show all the available data from the PDG08 [7]. On that figure Δ*-data stops at *J* = 15/2, but one should also include *Δ*17/2-(3300), *Δ*19/2+(3500), *Δ*19/2+(3700), and *Δ*21/2-(4100) from [7]. So, the curve will be naturally continued and the theoretical explanation is needed. Our NRQM high-*J* predictions gave a hint for the parity doublers

*Δ*17/2-(3300) [PDG]  -  *Δ*17/2+(3490) [2,3]

*Δ*19/2+(3700) [PDG]  -  *Δ*19/2-(4000) [2,3]

If one combines this with *N*\* higher –*J* states, the new chiral quartets arise [8]. Another good idea is to use our results [2-3] for the higher-*J N - Δ* spectra and RT to clarify completely claims [1]. For the first time *N-Δ* -spectra were computed up to the *L*=20, and appropriate RT's has been drawn and analyzed [3]. We are planning to complete the work and analyze highest possible parity doublets, Δ - quartets, and new *N - Δ* triplets. If we look at Fig.1 [3], u-d spectra, we see that $M^2(L=20) = 25$ GeV$^2$, or $M = 5$ GeV for the parent RT. Clearly, if we will extrapolate from the [1], Fig.3 spectra, it should become *flatten out* at such a high *J* ≥ 20. The spectra will become almost constant at high-*J*, and it means, there is a different scaling scenario taking place. Stepanov et al. [9] have got a similar boundary values for the charmonium, $M_{max} \approx 5.2$ GeV, $L_{max} \approx 12$. One can also witness from our Fig.5, [3] how the *Δ*\*'s RT curvature is dramatically changing over the whole *J*-interval, if we are accounting for higher *J* -states from the PDG [7]. So, it seems to be that a new regime started somewhere between 3 - 5 GeV in *N-Δ* mass, there are also different models with 3 or 4 different scales exist, and there is some difference between mesons and baryons physics [6].

On a page 10 the authors [1] wrote: "the degeneracy can be claimed for the 9/2 states alone, and the chiral partners higher in the spectrum are not experimentally known". In my recent paper [8], I found a lot of *Δ\** (and *N\**) parity doublers to be present in the whole *J*-region (see also recent review by R.Jaffe [10]). It is unclear, why there is such a discrepancy between theory and the data on Fig.3, [1] for *P*=+, *J*= 5/2, 7/2, 11/2, 13/2, 15/2 – it is about 500 MeV.

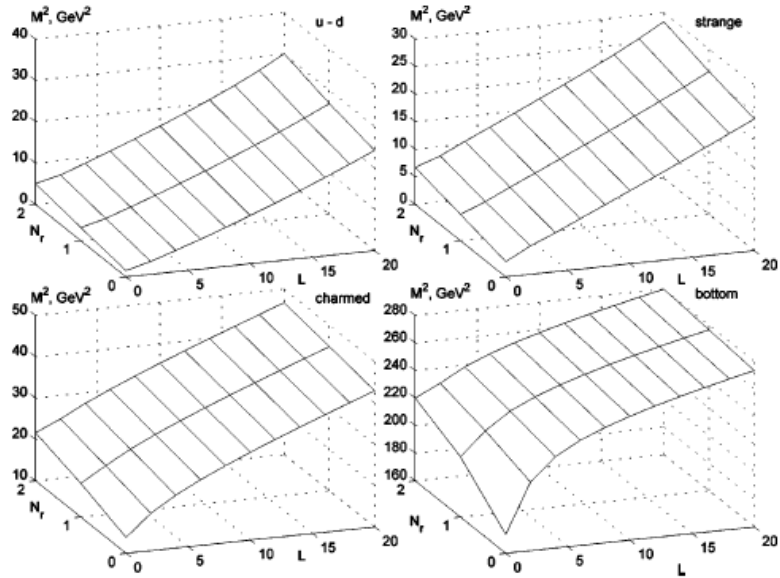

FIG. 1. Parent and two daughters RT in the potential quark model.

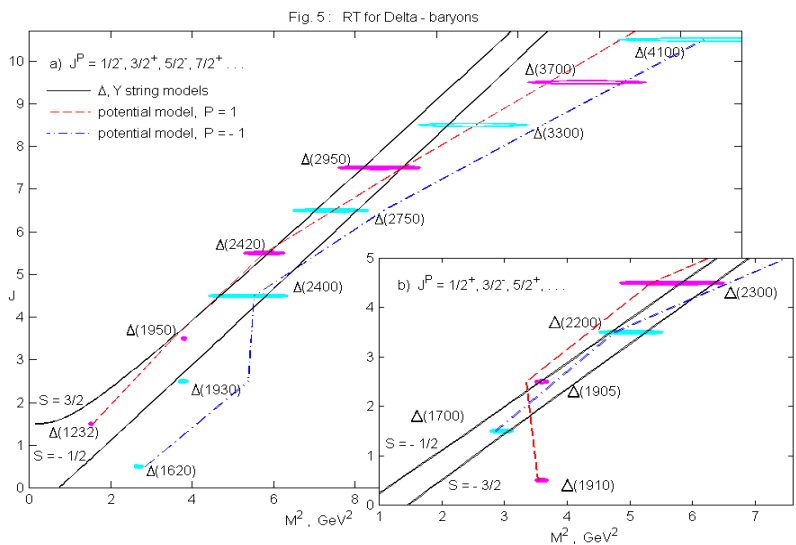

I wonder, how one should expect to get a reasonable *Δ\**- spectrum, rescaling pion wavefunction, which describes very compact, quasi-Goldstone particle (see p.11 [1])? Now, we reach the last Fig.4 in [1]. On the left there is splitting inside the first parity doublers versus *J*. One can immediately see, how huge are the splittings in model [1]. I'm not sure, whether authors confused some numbers and letters, but the interdoublet splittings of the order of 0.8 – 0.35 GeV are simply unacceptable and will ruin the notion of parity doublers from the start. The same story is for the right plot – interdoublet splittings between natural and unnatural parity states are also huge, with small reasonable subset for the unnatural parity. I think this failure is connected with high-momentum's truncation scheme, developed in [1], which is not justified for the modern $\Delta^*$ data from the PDG08 [7].

A. E. Inopin
Department of the Experimental Nuclear Physics
Karazin National University,
Freedom Square 4, Kharkov 61077, Ukraine